\documentclass[preprint,showpacs,preprintnumbers,amsmath,amssymb,prb]{revtex4} 
\usepackage{graphicx}   
\usepackage{epsfig}
\usepackage{latexsym}

\begin{document} 
\title{Thioglycolic acid on the gold (111) surface and Raman vibrational spectra}
\author{Jian-Ge Zhou$^{a}$, Quinton L. Williams$^{a}$, Ruqian Wu$^{b}$, }   
\affiliation{$^{a}$Department of Physics, Atmospheric Sciences, and Geoscience, Jackson 
State University, Jackson, MS 39217, USA\\
$^{b}$Department of Physics and Astronomy, University of California, Irvine, California 92697-4575, USA  }  
\author{}  
\affiliation{} 

\begin{abstract}
The interaction of thioglycolic acid (HSCH$_2$COOH) with the Au(111) 
surface is investigaged, and it is found that at the low coverage the molecule lies
down on the substrate. If the mercaptan-hydrogen atom is eliminated, the
resulting SCH$_2$COOH molecule is randomly oriented on the surface.
If the carboxylic acid group in the HSCH$_2$COOH molecule is deprotonated instead, the HSCH$_2$COO$^-$ molecule
lies down on the surface. However, when the mercaptan-hydrogen atom 
in the HSCH$_2$COO$^-$ molecule is removed, the resulting SCH$_2$COO$^-$ molecule
rises up to a certain level on the substrate. 
The calculated Raman vibrational spectra decipher which compounds and atomic displacements contribute
to the corresponding frequencies.
We thus propose a consistent
mechanism for the deposition of thioglycolic acid on the Au(111) surface. 

\end{abstract}  
\maketitle   

\section{Introduction}

Self-assembled monolayers (SAMs) have attracted considerable attention as 
model systems for many fundamental and technological investigations \cite{vvb}. The thiol and 
thiolate-based SAMs have broad applications on supramolecular assembly, wetting,
tribology, corrosion inhibition, lithography, chemical and biochemical sensors, optics,
and immobilization of DNA, because of both their simplicity and stability. In
particular, SAMs can simulate a biological membrane which allows  
adsorption of proteins to metal surfaces without denaturization \cite{pw1,beh,hhk}. 
The peptide molecules with some enzymatic activity can be deposited on metal
surfaces via the thiol or thiolate linkage monolayer \cite{mh,ah}. The deposition of a 
second monolayer on the top of the first adsorbed thiol or thiolate monolayer yields
a bilayer system consisting of two monomolecular films. In other words, a layer
of peptide molecules can be bonded to the gold surface via the linkage monolayer
formed from the thioglycolic acid (HSCH$_2$COOH) \cite{mh,ah,skm}. 

The chemisorption of the thioglycolic acid on the gold surface was demonstrated using surface-enhanced Raman
scattering \cite{kkm} and ultrafast electron crystallography \cite{ryz}. 
It was found when the higher portion of the carboxylic acid groups is deprotonated, the higher
portion of the thioglycolic acid molecules adopts a trans conformation \cite{kkm}.  
It was also observed that after 2,2'-dithiodiacetic acid is deposited on the Au(111),
the SCH$_2$COOH molecules are randomly oriented
on the gold surface \cite{ryz}, that is, the adsorption pattern related to 
the SCH$_2$COOH is different from that corresponding to the HSCH$_2$COOH. 
On the other hand, the swithcable SAM under the influence of an electrical 
potential was observed with intentionally created room for conformational changes of the
molecules \cite{lah}. When the external
electrical potential is turned on, the hexadecanoic acid molecules (HS(CH$_2$)$_{15}$COO$^-$)
bend their negatively charged COO$^-$ group
towards to the positively charged gold surface \cite{lah}.
Simulating this swithcable SAM process via the {\it ab initio} method 
requires a prohibitive amount of computer time, so one has to study the simple case: the 
HSCH$_2$COO$^-$ on the Au(111) surface.

It was recently observed that thiol stays intact when deposited on the regular
Au(111) surface, but the S-H bond
of the thiol is broken on the defected Au(111) surface \cite{rlm,zh}.
Upon the HSCH$_2$COOH molecules deposit on the Au(111), they can either remain intact,
or turn into one of the following substances: 1) SCH$_2$COOH in the presence of the
defect on the Au(111) \cite{rlm,zh}; 2) HSCH$_2$COO$^-$ by increasing pH value \cite{kkm};
3) SCH$_2$COO$^-$ by the defect and increasing pH value. 
To get a consistent picture of the thioglycolic acid adsorption on the 
Au(111), one has to examine the adsorption patterns of the HSCH$_2$COOH, 
SCH$_2$COOH, HSCH$_2$COO$^-$, and SCH$_2$COO$^-$ on the Au(111) separately.
Even some theoretical simulations on thiol or thiolate based SAMs have been carried out \cite{mbt,btp,gla,
gh,ghw,fp,zgy,alc}, however,
there has been no first-principle calculation which provides an atomic-scale 
description of the HSCH$_2$COOH, SCH$_2$COOH, HSCH$_2$COO$^-$, and SCH$_2$COO$^-$ on the Au(111) surface.
The electronic properties for this system, such as the projected
density of states (PDOS) and the charge density difference, have not been discussed. 
While a large variety of thiol or thiolate based SAMs has been studied, still little is known about 
why the SCH$_2$COOH molecules are randomly oriented on the gold surface, and
how the HSCH$_2$COOH molecules orient on the Au(111). 
Thus theory is challenged to propose a consistent model for the 
thioglycolic acid adsorption process on the Au(111) surface.

In this contribution, we address the adsorption patterns of the 
HSCH$_2$COOH, SCH$_2$COOH, HSCH$_2$COO$^-$ and SCH$_2$COO$^-$ molecules
on the Au(111) surface from first principle calculation. 
We present adsorption energies and geometries for these four kinds of molecules
on the Au(111) surface at 0.25 ML, and find that they demonstrate different adsorption patterns. 
We calculate the partial density of states (PDOS)
projected on the S and O2 atom (with an attached hydrogen, see Fig. \ref{structure}) 
to show their relation to the adsorption patterns, and 
evaluate the charge-density differences to illustrates the interacting bond between the adsorbates and
the Au(111). 
We also compute the Raman vibrational spectra of these four kinds of molecules adsorbed on the surface to 
decipher the adsorption mechanism of the thioglycolic acid on the Au(111) substrate.
By the comparison of the experimental frequencies with the computational ones,
we can identify which compounds and atomic displacements contribute to the corresponding frequencies.
We thus reveal how the dissociation of the mercaptan
hydrogen atom and the deprotonation of carboxylic acid group play key roles in the adsorption
process, and propose a consistent
mechanism for the deposition of thioglycolic acid on the Au(111) surface.

\section{Computational Method}

The calculations were carried out in the slab model with periodic boundary condtions by
density functional theory (DFT) \cite{khf,khf1}. The electron-ion interaction has been described using the projector
augmented wave (PAW) method \cite{kj,peb}.
All calculations have been performed by Perdew-Wang 91 (PW91) generalized gradient approximation \cite{pw}.
The wave functions were expanded in a plane wave basis with an energy cutoff of 400 eV. The $k$ points
were obtained from Monkhorst-Pack scheme \cite{mp1}, and $3\times 3\times 1$  $k$ point mesh was 
for the geometry optimization.
The optimization of the atomic geometry was performed via conjugate-gradient minimization of the
total energy with respect to the atomic coordinates.
The supercell consisted of five layers with each layer having 12 Au atoms.
The Au atoms in the top three atomic layers are allowed to relax, 
while those in the bottom two layers are fixed to simulate bulk-like termination \cite{zwh}. 
The vacuum region comprises ten atomic layers, which
exceeds substantially the extension of the thioglycolic acid molecule \cite{khf}.
For charged systems, a uniform compensating background is incorporated to maintain the charge neutrality of the 
supercell \cite{mp}. 
The harmonic approximation was applied to calculate the Hessian matrix and vibrational frequencies.
We calculated the gold lattice constant and found it to agree with the experimental value \cite{ak} 
within 2.1$\%$.

\section{Results and Discussion}
In this section, we discuss the adsorption pattern of the HSCH$_2$COOH, 
SCH$_2$COOH, HSCH$_2$COO$^-$, and SCH$_2$COO$^-$ on the Au(111) substrate, respectively.
The adsorption energy of the system is defined as $E_{ads}$ = $E_{adsorbate}$ + $E_{Au(111)}$ - $E_{adsorbate+Au(111)}$.
The symbol top-fcc (or top-hcp) in the following tables represents the S atom 
being on the atop site of the gold atom, but leaned toward the fcc (or hcp) hollow center,
and anologously for the notations bri-fcc, bri-hcp, etc. 
The units for the bond length and adsorption energy are Angstrom ($\AA$) and eV.

\subsection{The HSCH$_2$COOH molecule on the Au(111) surface}
First, let us begin with our analysis with the geometries and adsorption 
energies of the optimized structures for the thioglycolic acid on the Au(111) surface  
at the coverage of 0.25 ML, as displayed 
in Table \ref{cooh}. 
Here 1.00 ML  means one sulfur per three gold atoms, and 0.25ML stands for one 
thioglycolic acid  on a gold surface with twelve gold atoms. 
In Table \ref{cooh}, the entries $\theta$, $tilt$ $direct$ and $d_{S-Au}$  refer to
the polar angle between the normal vector of the surface and the S-C2 direction, the Au(111) surface region 
towards which the S-C2 is tilted, and the shortest Au-S bond length, respectively. 
The entries $initial$ and $optimized$ $site$ stand for
the S atom attachment site before and after optimization. The columns 1-3 and 4-7 list structural data 
pertaining to the initial
and the final optimized geometry. The maximum adsorption energy is underlined. 
\begin{table}
\caption{The geometries and adsorption energies for the structures 
of thioglycolic acid on Au(111) at 0.25ML. The entries $\theta$, $tilt$ $direct$ and $d_{S-Au}$  refer to
the polar angle between the normal vector of the surface and the S-C2 direction, the Au(111) surface region 
towards which
the S-C2 is tilted, and the shortest Au-S bond length. The entries $initial$ and $optimized$ $site$ stand for
the S atom attachment site before and after optimization. The maximum adsorption energy is underlined.}
\begin{center}
\begin{tabular}{cccccccc}
\hline
initial & $\theta$ & $d_{S-Au}$ & optimized & $\theta$ & tilt     & $d_{S-Au}$ & $E_{ads}$\\
site    & ~        & $\AA$       & site      & deg       & direct & $\AA$      & eV\\
\hline
  bri&0&2.60&bri&2.9&hcp&2.97&0.27\\
  ~&45&2.60&bri-fcc&54.8&hcp&2.89&0.45\\
  ~&90&2.60&bri&86.9&hcp&3.16&0.47\\
\hline
  fcc&0&2.60&bri-fcc&9.7&hcp&2.98&0.32\\
  ~&45&2.60&bri-fcc&61.0&hcp&3.11&0.36\\
  ~&90&2.60&top-fcc&83.7&hcp&3.18&0.50\\
\hline
  hcp&0&2.60&bri-hcp&5.8&fcc&2.95&0.30\\
  ~&45&2.60&bri-hcp&54.2&fcc&2.87&0.48\\
  ~&90&2.60&hcp&89.4&fcc&3.58&0.44\\
\hline
  top&0&2.60&top&8.0&hcp&2.93&0.26\\
  ~&45&2.60&top&54.8&hcp&2.81&0.37\\
  ~&90&2.60&top&74.2&fcc&2.57&$\underline{0.63}$\\
\hline
\end{tabular}
\end{center}
\label{cooh}
\end{table}

\begin{figure}
\begin{center}
\includegraphics[width=4.5in]{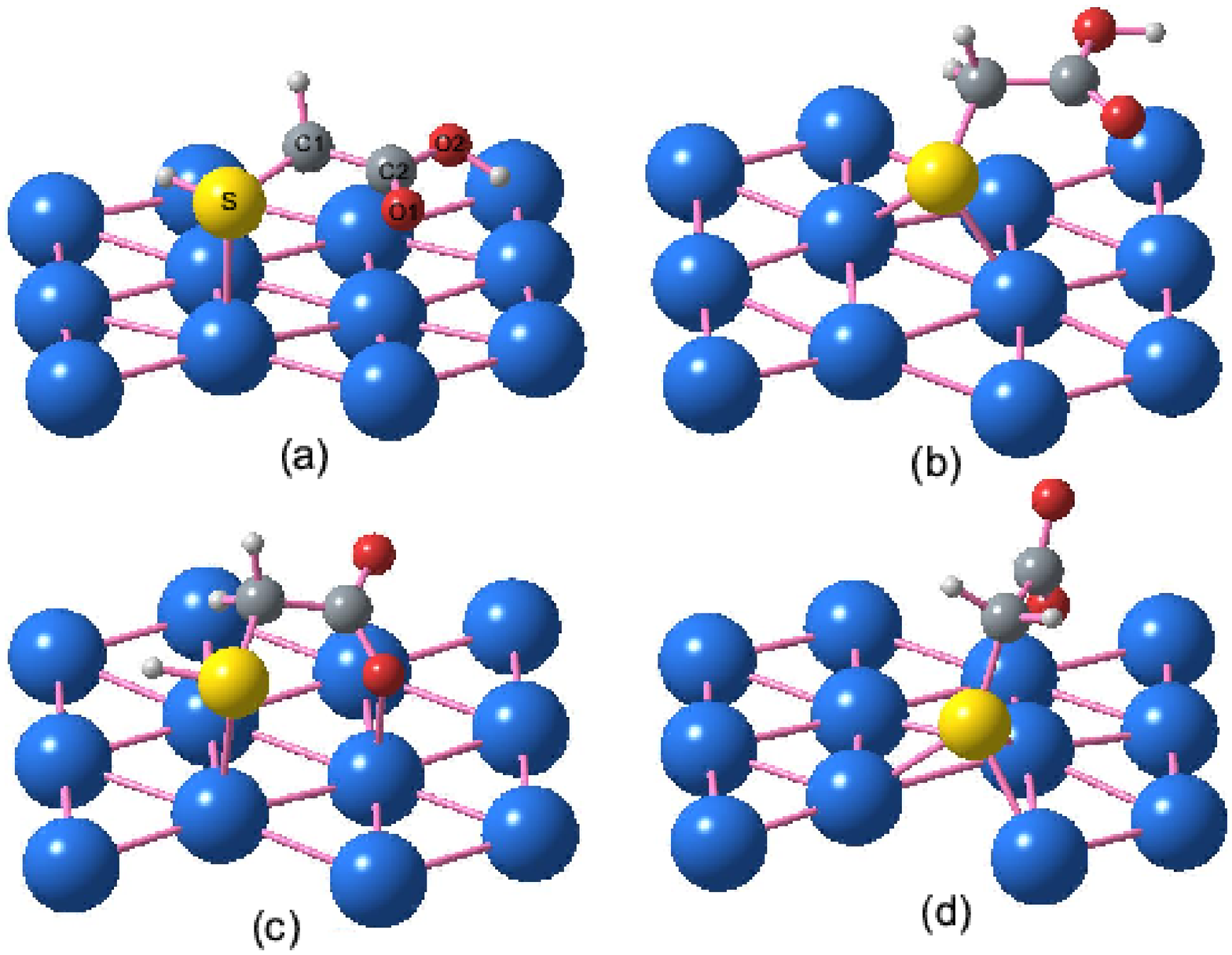}
\end{center}
\caption{(a) The thioglycolic acid (HSCH$_2$COOH) on the Au(111) surface. 
(b) SCH$_2$COOH on the surface.
(c) HSCH$_2$COO$^-$ on the surface. 
(d) SCH$_2$COO$^-$ on the surface.} 
\label{structure}
\end{figure}

Table \ref{cooh} shows the adsorption energy for the most stable structure of 
the HSCH$_2$COOH  on the Au(111) surface is 0.63 eV, and the adsorption site preferred
by the sulfur atom is located at the atop site of the gold atom. 
This stable configuration is illustrated in Fig. \ref{structure}a.
The polar angle between the normal vector of the surface and the S-C2 direction
is 74.2$^{\circ}$. Fig. \ref{structure}a indicates that at the low coverage, the HSCH$_2$COOH
tends to lie down \cite{kkm}. The S-Au bond length is 2.57 $\AA$, which suggests that the bonding between the S atom in the HSCH$_2$COOH  
and the gold atom could be described as chemisorption \cite{rlm}. 

\subsection{The SCH$_2$COOH molecule on the gold substrate}
Ruan et al. used 2,2'-dithiodiacetic acid to 
make SAMs on the gold surface \cite{ryz}, then the SCH$_2$COOH is deposited on the surface.
The SCH$_2$COOH can also be obtained from the HSCH$_2$COOH molecule by taking away the mercaptan hydrogen atom 
which is initially attached to the sulfur.
The adsorption pattern of the SCH$_2$COOH molecule 
on the Au(111) surface is depicted in Table \ref{cooh-late}. 
\begin{table}
\caption{The geometries and adsorption energies for the structures 
of the SCH$_2$COOH molecule on the Au(111) at 0.25ML.}
\begin{center}
\begin{tabular}{cccccccc}
\hline
initial & $\theta$ & $d_{S-Au}$ & optimized & $\theta$ & tilt     & $d_{S-Au}$ & $E_{ads}$\\
site    & ~        & $\AA$       & site      & deg       & direct & $\AA$      & eV\\
\hline
  bri&0&2.60&bri-hcp&6.5&fcc&2.48&2.26\\
  ~&$\#$45&2.60&bri-fcc&46.1&hcp&2.45&2.33\\
  ~&90&2.60&bri-fcc&71.3&hcp&2.42&2.17\\
\hline
  fcc&0&2.60&fcc&9.8&hcp&2.47&2.28\\
  ~&45&2.60&fcc&51.0&hcp&2.51&2.17\\
  ~&90&2.60&bri-fcc&71.3&hcp&2.50&2.23\\
\hline
  hcp&0&2.60&bri-hcp&2.7&fcc&2.47&2.22\\
  ~&$\#$45&2.60&bri-hcp&41.6&fcc&2.46&2.33\\
  ~&90&2.60&bri-hcp&68.0&fcc&2.47&$\underline{2.34}$\\
\hline
  top&0&2.60&top-fcc&10.1&hcp&2.40&2.01\\
  ~&45&2.60&top-fcc&63.0&hcp&2.40&1.89\\
  ~&90&2.60&top-fcc&70.3&hcp&2.59&2.05\\
\hline
\end{tabular}
\end{center}
\label{cooh-late}
\end{table}
The adsorption energy for the most stable configuration of 
the SCH$_2$COOH on the Au(111) surface in Table \ref{cooh-late} is 2.34 eV, and the favored adsorption site 
by the sulfur atom is in the hcp hollow center, but leaned to the Au-Au bridge. 
The corresponding structure is shown in Fig. \ref{structure}b.
The sulfur atom forms bonds with two Au atoms, and the S-Au bond length is
2.47 $\AA$. The angle between the normal vector of the surface and the S-C2 direction
is 68.0$^{\circ}$. However, in Table \ref{cooh-late} there are two configurations marked by $\#$ (bri-45$^{\circ}$
and hcp-45$^{\circ}$) with adsorption energies 2.33 eV, which is close to the energy
of the most stable one (2.34 eV). The corresponding angles $\theta$ for these two  
$\#$ marked configurations are
46.1$^{\circ}$ and 41.6$^{\circ}$, respectively.  This can be interpreted as when 
the SCH$_2$COOH molecules are deposited on the Au(111) surface, some SCH$_2$COOH molecules 
lie on the substrate (see Fig. \ref{structure}b), 
but some of them rise up
to a certain level (corresponding to $\#$ marked configurations). Thus in the case
of SCH$_2$COOH, the configurations with different tilted angles may admix, and the molecule
appears to deposit on the gold substrate randomly \cite{ryz}.

\subsection{The HSCH$_2$COO$^-$  molecule on the surface}
When the thioglycolic acid is adsorbed on the Au(111) surface, its carboxylic acid group (COOH) 
can be deprotonated and it becomes HSCH$_2$COO$^-$ \cite{kkm}.
The optimized adsorption configurations of the HSCH$_2$COO$^-$ molecule on the Au(111) surface
are described in Table \ref{coo}.  
\begin{table}
\caption{The geometries and adsorption energies for the structures 
of the HSCH$_2$COO$^-$ molecule on the Au(111) surface at 0.25ML.}
\begin{center}
\begin{tabular}{cccccccc}
\hline
initial & $\theta$ & $d_{S-Au}$ & optimized & $\theta$ & tilt     & $d_{S-Au}$ & $E_{ads}$\\
site    & ~        & $\AA$       & site      & deg       & direct & $\AA$      & eV\\
\hline
  bri&0&2.60&bri&1.1&hcp&2.86&0.44\\
  ~&45&2.60&bri&50.6&hcp&2.66&0.67\\
  ~&90&2.60&bri&82.8&hcp&2.99&0.96\\
\hline
  fcc&0&2.60&bri-fcc&8.0&fcc&2.67&0.54\\
  ~&45&2.60&bri-fcc&65.4&fcc&2.74&0.92\\
  ~&90&2.60&top-fcc&77.1&hcp&3.38&0.95\\
\hline
  hcp&0&2.60&hcp&1.5&fcc&2.75&0.66\\
  ~&45&2.60&top-hcp&46.6&fcc&2.57&0.81\\
  ~&90&2.60&bri-hcp&78.0&fcc&3.62&0.83\\
\hline
  top&0&2.60&top&5.7&hcp&2.85&0.30\\
  ~&45&2.60&top&56.6&hcp&2.65&0.94\\
  ~&90&2.60&top&82.9&hcp&2.60&$\underline{1.13}$\\
\hline
\end{tabular}
\end{center}
\label{coo}
\end{table}
The adsorption energies in Table \ref{coo} demonstrate that the sulfur atom in
the HSCH$_2$COO$^-$ molecule prefers to stay
on the atop site of the gold atom, as indicated in Fig. \ref{structure}c. 
The corresponding adsorption energy is 1.13 eV which is larger than that
of the thioglycolic acid on the Au(111) surface. The S-Au bond length is around
2.60 $\AA$ and the angle $\theta$ is   
82.9$^{\circ}$, which means the HSCH$_2$COO$^-$ molecule is lying down on the gold substrate.

\subsection{The SCH$_2$COO$^-$ molecule on the Au(111) surface}
If the mercaptan-H atom in the HSCH$_2$COO$^-$ is detached from the sulfur atom,
the resulting compound is the SCH$_2$COO$^-$.  
Table \ref{coo-late} shows that the sulfur atom favors the fcc hollow center with the
adsorption energy 2.34 eV. The polar angle between the normal vector of the 
gold surface and the S-C2 direction is 53.4$^{\circ}$; so, after losing mercaptan-H atom,
the SCH$_2$COO$^-$ molecules begin to rise, see
Fig. \ref{structure}d. Note that
there is a $\#$ marked configuration in Table \ref{coo-late}, whose adsorption
energy is 2.33 eV - very closed to 2.34 eV. The angle $\theta$ for top-45$^{\circ}$ structure (the
$\#$ marked configuration) is 51.8$^{\circ}$.  The stable structures in Table \ref{coo-late}
indicate that when  the SCH$_2$COO$^-$ is deposited on
the Au(111) surface, the molecule rises up to a certain level \cite{kkm} (see Fig. \ref{structure}d). 
\begin{table}
\caption{The geometries and adsorption energies for the structures 
of the SCH$_2$COO$^-$ molecule on the Au(111) surface at 0.25ML.}
\begin{center}
\begin{tabular}{cccccccc}
\hline
initial & $\theta$ & $d_{S-Au}$ & optimized & $\theta$ & tilt     & $d_{S-Au}$ & $E_{ads}$\\
site    & ~        & $\AA$       & site      & deg       & direct & $\AA$      & eV\\
\hline
  bri&0&2.60&bri-hcp&9.2&fcc&2.51&2.16\\
  ~&45&2.60&bri&44.9&hcp&2.45&2.01\\
  ~&90&2.60&bri-fcc&78.6&hcp&2.43&2.21\\
\hline
  fcc&0&2.60&fcc&7.8&hcp&2.47&2.14\\
  ~&45&2.60&bri-fcc&53.4&hcp&2.49&$\underline{2.34}$\\
  ~&90&2.60&bri-fcc&83.2&hcp&2.49&2.28\\
\hline
  hcp&0&2.60&hcp&0.5&fcc&2.49&2.11\\
  ~&45&2.60&bri-hcp&39.8&fcc&2.44&2.04\\
  ~&90&2.60&bri-hcp&76.4&fcc&2.45&2.27\\
\hline
  top&0&2.60&bri-top&10.1&bri&2.54&1.88\\
  ~&$\#$45&2.60&bri-hcp&51.8&fcc&2.43&2.33\\
  ~&90&2.60&top&81.8&fcc&2.38&2.08\\
\hline
\end{tabular}
\end{center}
\label{coo-late}
\end{table}

\subsection{Electronic Structures}
\begin{figure}
\begin{center}
\includegraphics[width=5.9in]{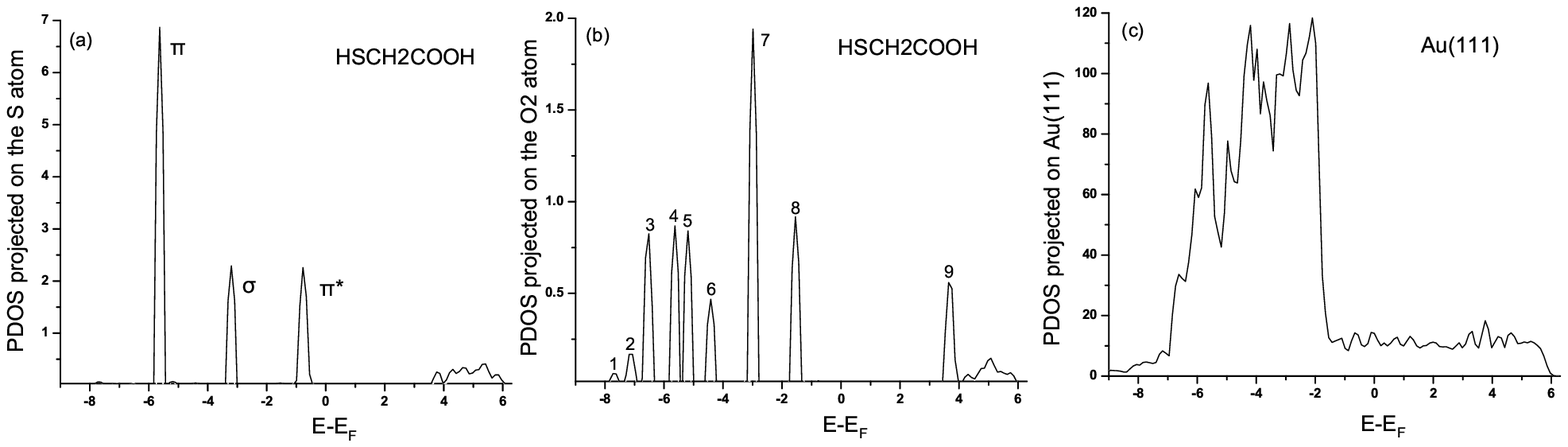}
\end{center}
\caption{(a)-(b) PDOS projected on the S and O2 atom in the isolated HSCH$_2$COOH,
(c) DOS for a pure Au(111) surface.} 
\label{dos1}
\end{figure}

\begin{figure}
\begin{center}
\includegraphics[width=6.5in]{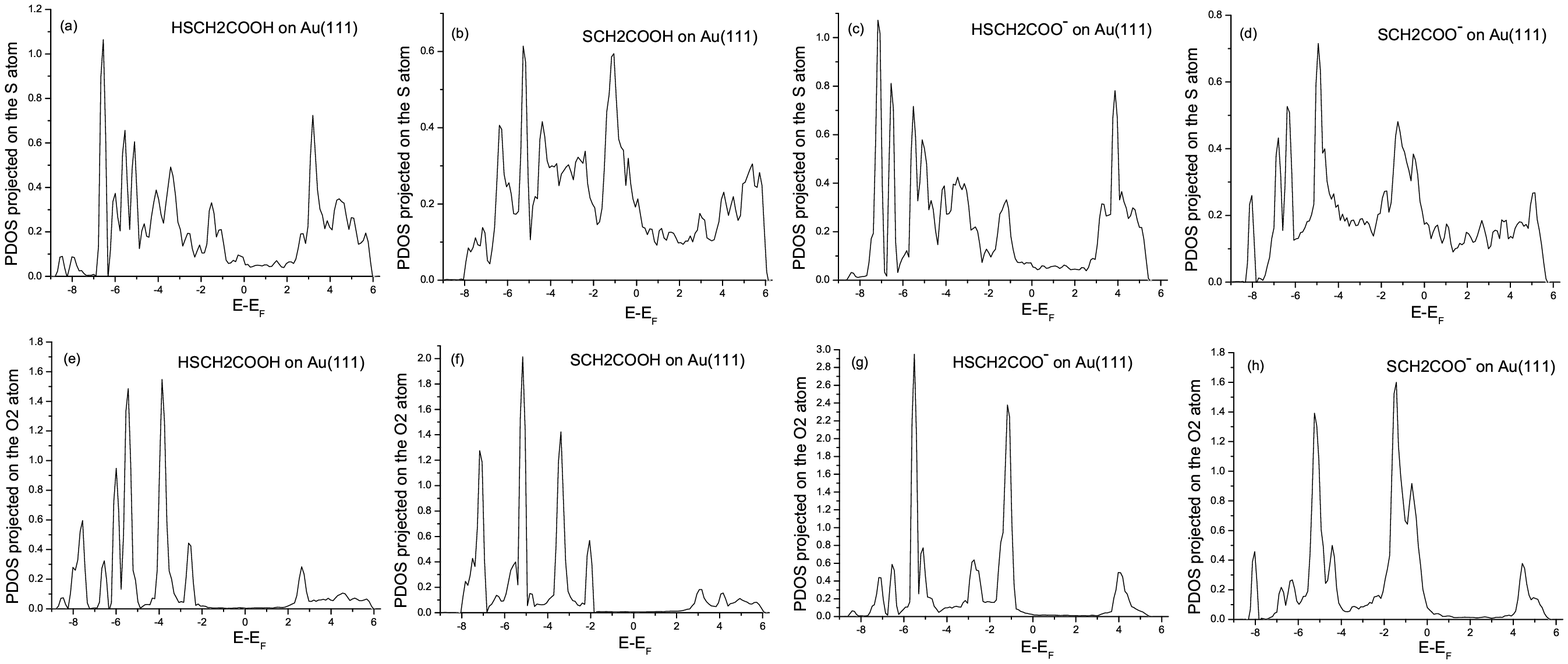}
\end{center}
\caption{(a)-(d) PDOS on the S in the HSCH$_2$COOH, SCH$_2$COOH, HSCH$_2$COO$^-$ and SCH$_2$COO$^-$ on the surface,  
(e)-(h) PDOS on the O2 in the HSCH$_2$COOH, SCH$_2$COOH, HSCH$_2$COO$^-$ and SCH$_2$COO$^-$ on the surface.} 
\label{dos2}
\end{figure}

To understand how the dissociation of the mercaptan-hydrogen atom and 
the deprotonation of the carboxylic acid group play roles in the adsorption
process, we calculate the partial density of states (PDOS)
projected on the S and O2 atoms in the HSCH$_2$COOH, SCH$_2$COOH, 
HSCH$_2$COO$^-$, SCH$_2$COO$^-$ molecule deposited on the Au(111) substrate. 
There are three sharp peaks in the PDOS projected on the
S atom in the isolated HSCH$_2$COOH molecule (Fig. \ref{dos1}a). 
The major contributions of three
peaks come from $\pi$, $\sigma$ and $\pi^*$ orbitals in the S-C bond. 
To calibrate the Fermi level for the isolated molecule, in the calculation we have seperated
the HSCH$_2$COOH molecule from the gold surface by 8$\AA$ so that there is no interaction 
between the molecule and the substrate. The corresponding PDOS projected on the S atom
can be regarded as that in the isolated HSCH$_2$COOH molecule. The density of states for the
pure gold surface vanishes above 6eV (Fig. \ref{dos1}c). The $\pi^*$ 
orbital is located on the right edge of the Au d band, the $\pi$ orbital is near the left
edge of the Au d band, whereas the $\sigma$ orbital overlaps with the gold d band (Fig. \ref{dos1}a,
Fig. \ref{dos1}c). 
Upon the HSCH$_2$COOH molecule is deposited on the surface (Fig. \ref{dos2}a),
the $\sigma$ and $\pi^*$ states disperse as a consequence of the mixing with the gold
d states, whereas $\pi$ orbital remains sharp and shifts toward more negative energy. 
The adsorption energy of 0.63eV indicates that the major interaction between the sulfur in  the HSCH$_2$COOH molecule 
and the gold surface is not van der Waals force, so  Fig. \ref{dos2}a might not have the signature of the weak bond.

If the mercaptan-hydrogen atom is dissociated from the S atom, the $\pi$ state splits indicating
a stronger S-Au bonding interaction originating from the hybridization of the $\pi$ orbital
of the SCH$_2$COOH with the gold d band (Fig. \ref{dos2}b).  The PDOS projected on the S 
atom is insensitive to the deprotonation of the carboxylic acid group, which explains why 
the profiles of Fig. \ref{dos2}a and  Fig. \ref{dos2}c are similar so are Fig. \ref{dos2}b and
Fig. \ref{dos2}d. No energy gap in the PDOS projected on the S atom 
attached to the Au(111) surface (Fig. \ref{dos2}a-\ref{dos2}d). This is because the HOMO and LUMO level of the S atom
fall into the energy range of the gold d-band with a concomitant hybridization, that is, the PDOS projected on the S atom 
near the Fermi level is dominated by d-states from the Au(111) surface.
The effect of the deprotonation of the carboxylic acid group is demonstrated
by the PDOS projected on the O2 atom (Fig. \ref{dos2}e-Fig. \ref{dos2}h).
Nine sharp peaks are illustrated in the PDOS projected on the O2 atom in the isolated HSCH$_2$COOH
molecule (Fig. \ref{dos1}b). Upon the HSCH$_2$COOH is deposited on the gold surface, some
peaks are suppressed (Fig. \ref{dos2}e), however, the PDOS projected on the O2 atom
is insensitive to the adsorption and the dissociation of the mercaptan-hydrogen atom 
(Fig. \ref{dos2}e, Fig. \ref{dos2}f). Fig. \ref{dos2}g and Fig. \ref{dos2}h
show that some peaks in Fig. \ref{dos1}b disappear and
some others disperse after the carboxylic acid group is deprotonated, which indicates that the deprotonation
changes the electronic states around the O2 atom. 

\begin{figure}
\begin{center}
\includegraphics[width=4.5in]{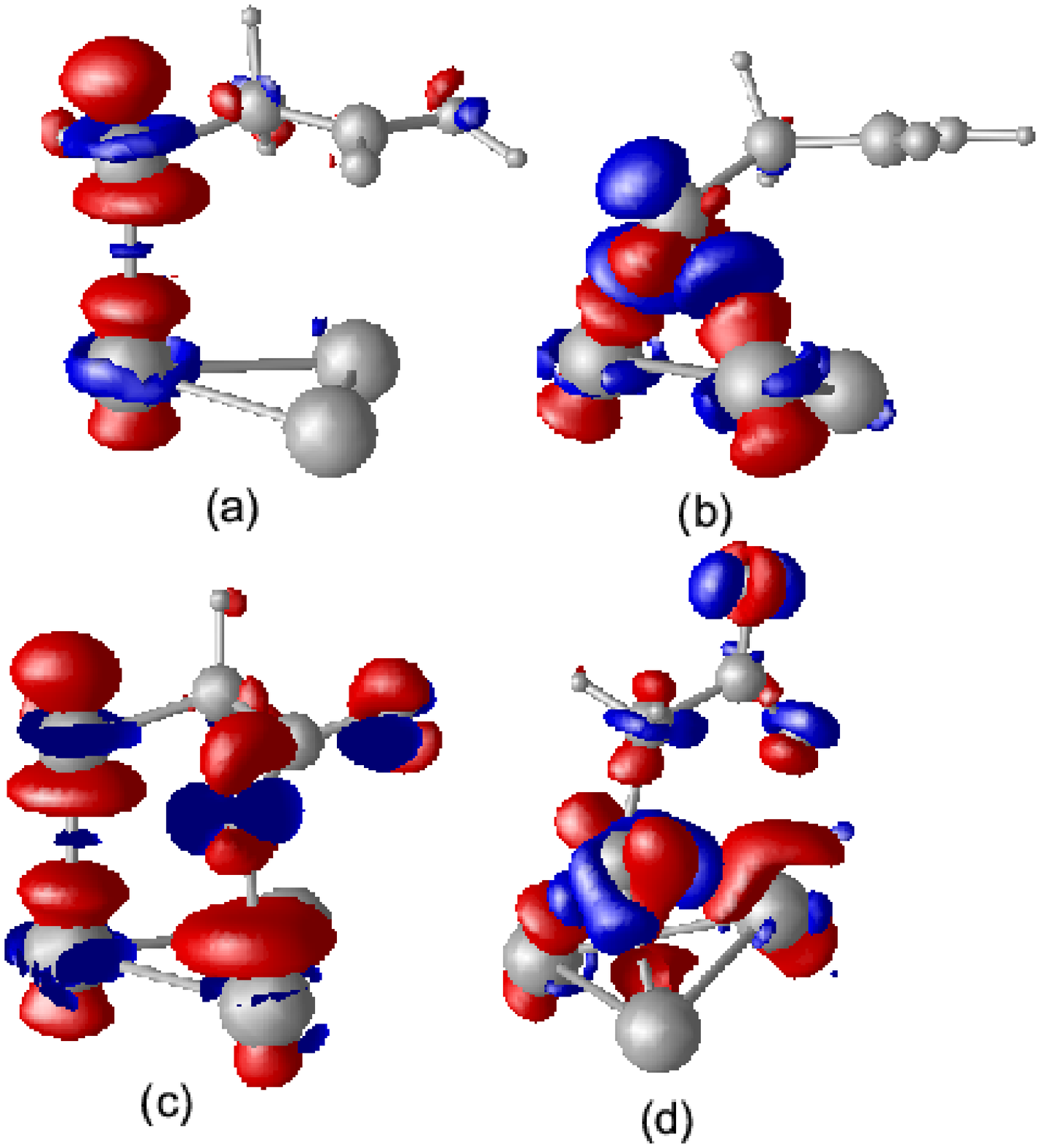}
\end{center}
\caption{The isosurfaces of the
charge-density difference for (a) the HSCH$_2$COOH adsorption
on the Au(111) surface with blue (accumulation of electrons) and/or red (depletion
of electrons) isosurface value, ${\pm}0.02e/{\AA^{3}}$,  
(b) SCH$_2$COOH on the surface, 
(c) HSCH$_2$COO$^-$ on the surface, and
(d) SCH$_2$COO$^-$ on the surface.
Only three related gold atoms of the Au(111) surface are displayed.} 
\label{chgdiff}
\end{figure}

To further elucidate the interacting bond between the HSCH$_2$COOH, SCH$_2$COOH, 
HSCH$_2$COO$^-$, SCH$_2$COO$^-$ molecule and
the Au(111) substrate, we calculate the charge-density difference:
$\Delta\rho(\vec{r})=\rho_{ads/sub}(\vec{r})-\rho_{sub}(\vec{r})-\rho_{ads}(\vec{r})$,
where $\rho_{ads/sub}$, $\rho_{sub}$, and $\rho_{ads}$ are the electron charge densities
of the relaxed adsorbate-substrate system, of the clean relaxed surface,
and of the isolated but adsorptionlike deformed adsorbate (without substrate), respectively.
The isodensity surfaces of the charge-density difference for the 
structures of the HSCH$_2$COOH, SCH$_2$COOH, HSCH$_2$COO$^-$, and SCH$_2$COO$^-$ on 
the Au(111) substrate
are depicted in Figs. \ref{chgdiff}a-\ref{chgdiff}d. In Fig. \ref{chgdiff}, 
we display only the surrounding part of the S-Au bond. 
As we know, two p-electrons of the sulfur in the HSCH$_2$COOH molecule form a lone pair, and the region around the top of the gold is  
a charge depletion area \cite{vd}. In the configuration of Fig. \ref{chgdiff}a, the sulfur atom sits on the top of the gold atom. 
The lone pair in the sulfur is attracted to this charge depletion region, and 0.3e is transferred from 
the sulfur lone pair orbital to the gold charge depletion area.
Thus when the sulfur atom in the thioglycolic acid is adsorbed
on the Au(111) (Fig. \ref{chgdiff}a), the
electrostatic interaction responsible for the bonding comes from the 
monopole term and the dipole moments in the adsorbate and substrate \cite{psb}. Around
the sulfur atom, there is a ``ring" of accumulation of electron charge. 
The electrostatic interaction is dominated by the attractive ionic term 
modified by a repulsive dipolar term \cite{psb}.  
The sulfur atom stays on the top of the 
gold atom, that is, the S atom only forms a bond with one gold atom.
If the mercaptan-hydrogen atom is detached from the sulfur (Fig. \ref{chgdiff}b), the S-Au 
bond is largely covalent with some ionic character \cite{eu}. 
The sulfur in the SCH$_2$COOH forms bonds with two gold atoms of the Au(111) surface. 
The interaction between the HSCH$_2$COO$^-$
and the gold surface is the similar to the thioglycolic acid case (Fig. \ref{chgdiff}c),
except that the O1 atom form an additional bond with the gold atom.
There is big depletion of electrons around the gold atom. The $p_x$, $p_y$
orbits in the O1 atom gain extra electrons, but the $p_z$ orbit loses
some electrons. Thus, this O1-Au bonding is a convolution between the ionic bond
and the covalent bond. Fig. \ref{chgdiff}d suggests that the S-Au bonds for  
the SCH$_2$COO$^-$ is a covalent bond with some ionic character.

\subsection{The adsorption mechanism}
From the above discussion, we propose the following picture. When the HSCH$_2$COOH molecule is
adsorbed on the gold substrate, it lies down on the surface \cite{kkm} (Fig. \ref{structure}a). 
When the 2,2'-dithiodiacetic acid 
is put on the gold surface, the SCH$_2$COOH molecules form the SAM on the substrate \cite{ryz}. 
If the thioglycolic acid is deposited on the gold surface with defects, 
the mercaptan hydrogen atom can be dissociated
from the S atom and the HSCH$_2$COOH molecule becomes SCH$_2$COOH \cite{rlm,zh}. 
Some SCH$_2$COOH molecules lie down on the substrate (see Fig. \ref{structure}b), 
but others rise up to a certain level. Thus, in the case
of SCH$_2$COOH, different configurations may admix and the adsorption
appears to be randomly oriented \cite{ryz}.
If the carboxylic acid group in the HSCH$_2$COOH molecule is deprotonated by
increasing the pH value, the resulting HSCH$_2$COO$^-$ lies on the surface (Fig. \ref{structure}c).
However, when the mercaptan hydrogen atom in the HSCH$_2$COO$^-$ molecule
is ruptured from the sulfur, the resulting SCH$_2$COO$^-$ molecule rises up to a certain level \cite{kkm} (Fig. \ref{structure}d). 

\section{The Raman vibrational spectra}
To support the above adsorption mechanism, 
we calculate the Raman vibrational spectra of the HSCH$_2$COOH, SCH$_2$COOH, 
HSCH$_2$COO$^-$ and SCH$_2$COO$^-$ adsorbed on the Au(111) substrate, respectively.
The Raman vibrational peak frequencies (cm$^{-1}$) for experimental data and computational
values for the HSCH$_2$COOH, SCH$_2$COOH, HSCH$_2$COO$^-$ and SCH$_2$COO$^-$ on the Au(111) surface
are listed in Table \ref{freq}. The vibrational frequencies were calculated for the most stable configurations at 0.25ML.
\begin{table}
\caption{Raman vibrational peak frequencies (cm$^{-1}$): Experimental data and computational
values for the HSCH$_2$COOH, SCH$_2$COOH, HSCH$_2$COO$^-$ and SCH$_2$COO$^-$ on the Au(111) surface.
The calculated frequencies which are the closest to the experimental ones are underlined.}
\begin{center}
\begin{tabular}{ccccccccc}
\hline
frequencies & $\omega_{1}$ & $\omega_{2}$   & $\omega_{3}$  & $\omega_{4}$   & $\omega_{5}$    & $\omega_{6}$ & $\omega_{7}$ & $\omega_{8}$ \\
\hline
Exp. data \cite{kkm}& 575&665&763&905&930&1387&1597&1711\\
Au-HSCH$_2$COOH & \underline{575}&653&\underline{745}&\underline{881}&-&1409&-&\underline{1733}\\
Au-SCH$_2$COOH & 596&\underline{670}&725&872&-&\underline{1396}&-&1738\\
Au-HSCH$_2$COO$^-$ & 592 & 619&734&-&\underline{929}&1369&1627&-\\
Au-SCH$_2$COO$^-$& 598 & 638&822&-&838&1366&\underline{1577}&-\\
  \hline
\end{tabular}
\end{center}
\label{freq}
\end{table}
The calculated frequencies which are the closest to the experimental ones are underlined.
The Raman scattering is limited to the center of the Brillion zone, and
the vibrational frequencies are calculated at $\bar{\Gamma}$ point.

\begin{figure}
\begin{center}
\includegraphics[width=4.6in]{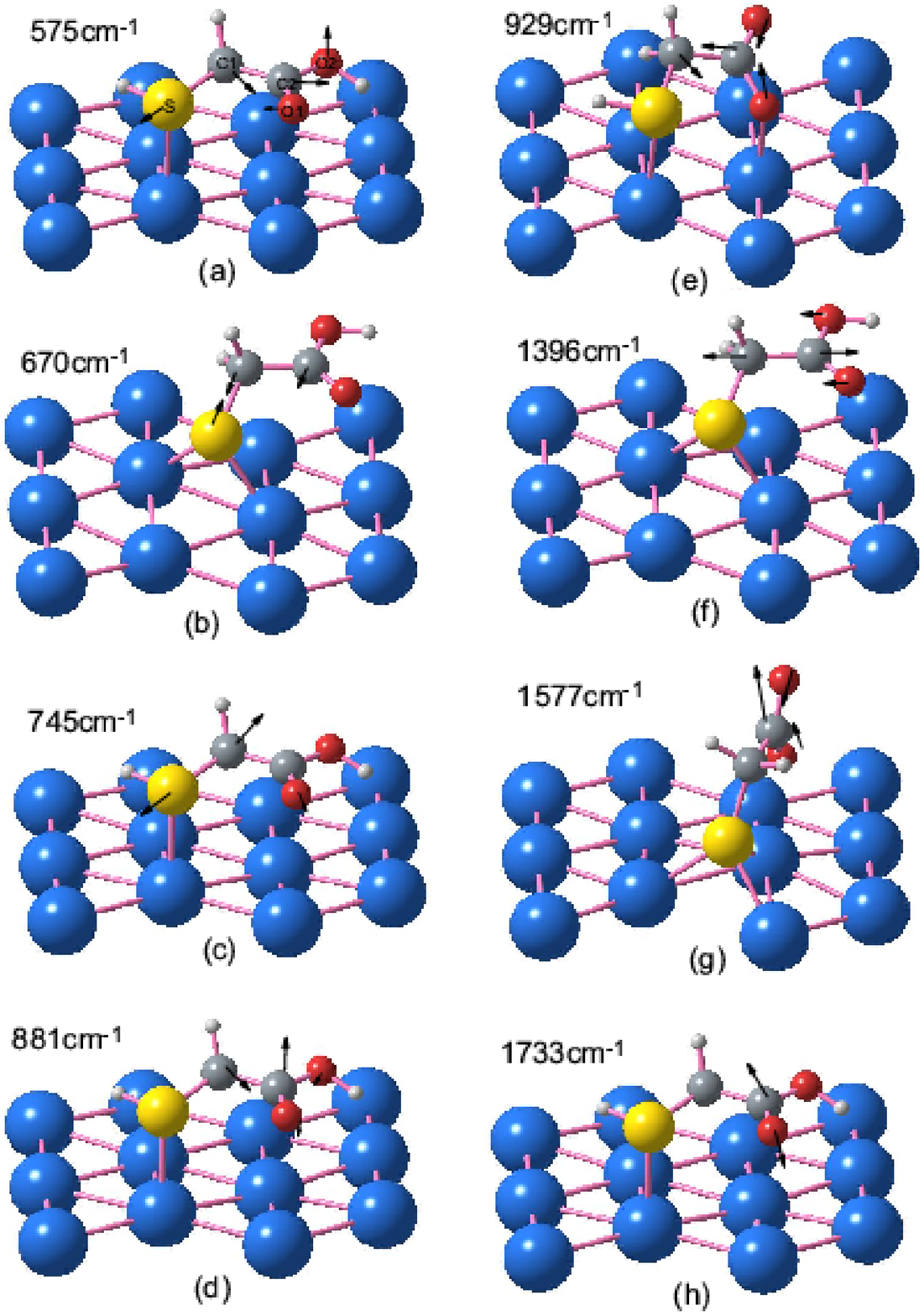}
\end{center}
\caption{(a)-(h) illustrate the calculated atomic displacements for the frequencies $\omega_{1}$ - $\omega_{8}$.} 
\label{disp}
\end{figure}

As shown in Fig. \ref{disp}a, the frequency $\omega_{1}$ (575cm$^{-1}$, 575cm$^{-1}$) (the first number stands for the 
experimentally measured frequency and the second one is the corresponding theoretical one,
which is underlined in Table \ref{freq}) is the vibration of the S-C1 and the C1-C2 
stretches (the O1 and O2 atoms displace slightly).
The theoretical counterpart suggests that the frequency 575cm$^{-1}$ comes from the HSCH$_2$COOH on the gold surface. 
Fig. \ref{disp}b indicates that the mode $\omega_{2}$ (665cm$^{-1}$, 670cm$^{-1}$) is attributed to
the C1-S vibration of the SCH$_2$COOH  on the substrate.
The frequency $\omega_{3}$ (763cm$^{-1}$, 745cm$^{-1}$) corresponds to the C1-S stretch of 
the HSCH$_2$COOH on the surface, where the O1 atom displaces slightly (Fig. \ref{disp}c).
The mode $\omega_{4}$  (905cm$^{-1}$, 881cm$^{-1}$) is ascribed to the stretching vibration
of the C-COOH in the HSCH$_2$COOH molecule on the Au(111) (Fig. \ref{disp}d). 
$\omega_{5}$ (930cm$^{-1}$, 929cm$^{-1}$) corresponds
to the C-COO$^-$ stretching vibration for the HSCH$_2$COO$^-$ on the gold surface (Fig. \ref{disp}e). The
frequency $\omega_{6}$ (1387cm$^{-1}$, 1396cm$^{-1}$) is due to the vibration
of the COOH of the SCH$_2$COOH on the gold surface (the C1 atom moves slightly,
see Fig. \ref{disp}f). The mode $\omega_{7}$ (1597cm$^{-1}$, 1577cm$^{-1}$)
can be assigned to this COO$^-$ stretch in the SCH$_2$COO$^-$ (Fig. \ref{disp}g), which indicates that
after the dissociation of the mercaptan
hydrogen atom and the deprotonation of the carboxylic acid group, 
some original HSCH$_2$COOH molecules on the surface have turned into the SCH$_2$COO$^-$ molecules. 
The frequency $\omega_{8}$ (1711cm$^{-1}$, 1733cm$^{-1}$) corresponds
to the C=O stretching vibration for the HSCH$_2$COOH on the Au(111) substrate (Fig. \ref{disp}h). 
Thus the above frequency comparison suggests that after the HSCH$_2$COOH molecules deposited
on the Au(111) surface, some stay intact on the surface, the rest have turned into 
SCH$_2$COOH (via dissociation), HSCH$_2$COO$^-$ (deprotonation) and 
SCH$_2$COO$^-$ (dissociation and deprotonation).

\section{Conclusion}

We have discussed the adsorption patterns of the HSCH$_2$COOH, 
SCH$_2$COOH, HSCH$_2$COO$^-$, and SCH$_2$COO$^-$ molecules on the Au(111) substrate by 
first-principle theoretical calculation. 
We have computed the partial density of states (PDOS)
projected on the S and O2 atoms in the molecules on
the Au(111) substrate, which display how the dissociation of the mercaptan
hydrogen atom and the deprotonation of the carboxylic acid group affect the deposition and 
the corresponding electronic configuration.  
We have calculated the charge-density differences for 
the molecules on the Au(111) substrate,
which illustrates various bonding characteristics. 
We have also studied the Raman vibrational spectra of the molecules adsorbed on the Au(111) substrate,
and by the comparison of the experimental frequencies with the computational ones,
we have identified which compounds and atomic displacements contribute to the frequencies. 
We have found the following adsorption mechanism for the thioglycolic acid on the Au(111) surface.
Upon the HSCH$_2$COOH molecules deposit on the Au(111), they can either remain intact,
or turn into one of the following substances: 1) SCH$_2$COOH in the presence of the
defect on the Au(111); 2) HSCH$_2$COO$^-$ by increasing pH value;
3) SCH$_2$COO$^-$ by the defect and increasing pH value. 
If the intact HSCH$_2$COOH is
adsorbed on the gold substrate, the molecule lies down on the surface. When the SCH$_2$COOH molecules deposited on the Au(111) surface,
some SCH$_2$COOH molecules lie on the substrate, 
but others rise up to a certain level. Thus, in the case
of SCH$_2$COOH, different configurations may admix, and the molecules
appear to be deposited on the gold substrate in a random fashion.
If the carboxylic acid group in the HSCH$_2$COOH is deprotonated, the resulting HSCH$_2$COO$^-$ lies down on the surface.
However, when the S-H bond in the HSCH$_2$COO$^-$ 
is broken and the molecule is turned into SCH$_2$COO$^-$, 
it rises up to a certain level.

\section*{Acknowledgments}

This work is funded in part by the DoD (Grant No. W912HZ-06-C-0057) and 
DOE-BES (Grant No. DE-FG02-04ER15611).

\end{document}